\documentclass[10pt,letterpaper]{article}
\usepackage{opex3}
\usepackage{amsmath}
\usepackage{amssymb}
\usepackage{braket}
\usepackage[labelsep=period]{caption,subcaption}
\usepackage{subfig}

\usepackage{soul}
\usepackage{ulem}
\usepackage{lineno}
\usepackage{bm}% bold math
\usepackage{paralist}
\usepackage[usenames,dvipsnames,svgnames,table]{xcolor}
\usepackage{titlesec}
\titleformat*{\subsubsection}{\normalsize}

\begin{document}
%% NOTE: TITLE PAGE & TOC NOT USED FOR MANUSCRIPT SUBMISSIONS %%
%\title{Photon Emission Rate Engineering using Graphene Nanodisc Cavities}

%\vskip4pc

%\tableofcontents
%\clearpage
%% NO TITLE PAGE FOR OPEX SUBMISSIONS %%

%% START HERE
%%%%%%%%%%%%%%%%%% title page information %%%%%%%%%%%%%%%%%%
\title{Photon emission rate engineering using graphene nanodisc cavities}

\author{Anshuman Kumar,$^{1}$ Kin Hung Fung,$^{2}$ M. T. Homer Reid,$^{3}$ and Nicholas X. Fang$^{1,*}$}

\address{$^1$ Mechanical Engineering Department, Massachusetts Institute of Technology, Cambridge, MA - 02139, USA\\ $^2$ Department of Applied Physics, The Hong Kong Polytechnic University, Hong Kong, China\\ $^3$ Mathematics Department, Massachusetts Institute of Technology, Cambridge, MA - 02139, USA
}
\email{*nicfang@mit.edu} %% email address is required
\begin{abstract}In this work, we present a systematic study of the plasmon modes in a system of vertically stacked pair of graphene discs. Quasistatic approximation is used to model the eigenmodes of the system. Eigen-response theory is employed to explain the spatial dependence of the coupling between the plasmon modes and a quantum emitter. These results show a good match between the semi-analytical calculation and full-wave simulations. Secondly, we have shown that it is possible to engineer the decay rates of a quantum emitter placed inside and near this cavity, using Fermi level tuning, via gate voltages and variation of emitter location and polarization. We highlighted that by coupling to the bright plasmon mode, the radiative efficiency of the emitter can be enhanced compared to the single graphene disc case, whereas the dark plasmon mode suppresses the radiative efficiency. \end{abstract}
 % REPLACE WITH CORRECT OCIS CODES FOR YOUR ARTICLE
%%%%%%%%%%%%%%%%%%%%%%% References %%%%%%%%%%%%%%%%%%%%%%%%%

%%%%%%%%%%%%%%%%%%%%%%%%%%  body  %%%%%%%%%%%%%%%%%%%%%%%%%%
%start line-numbering for internal editing purposes
%\linenumbers

\section{Introduction}

Technological advances in the field of nanofabrication have provided a powerful tool to tailor light-matter interaction. Metallic nanoparticles support surface plasmon resonances where collective oscillations of electron and photons can result in a localization of electromagnetic fields into subwavelength scales \cite{MaierBook}. Apart from localizing the incident plane waves, these plasmon modes strongly modify spontaneous emission properties of quantum emitters \cite{0022-3727-41-1-013001}, such as quantum dots, placed close to them. In particular, radiative decay rates of such fluorescent particles can be tuned, depending on whether the emitter couples to a radiative or non-radiative plasmon mode \cite{10.1103/PhysRevLett.102.107401}. This approach of tunable fluorescence quenching and enhancement finds its uses in applications such as molecular imaging \cite{doi:10.1021/nl902709m}. A number of geometries have been explored for such decay rate engineering, for instance metallic planar surface \cite{10.1063/1.431483}, photonic crystals \cite{10.1038/nphoton.2007.141} and various collections of metal nanoparticles \cite{10.1103/PhysRevLett.102.107401}. In particular, collections of nanoparticles, such as dimers provide an additional parameter, namely, separation and orientation of the individual particles with respect to each other to tune the local electromagnetic density of states. In certain dimer systems with inversion symmetry, symmetric or antisymmetric dipole modes can be excited, even using a plane wave, in accordance with the plasmon hybridization model. The antisymmetric or dark mode has a cancellation of the two induced electric dipole moments hence suppresses far field radiation. The converse is true for the symmetric or bright mode where the electric dipole moments add up constructively.

Metals, as plasmonic materials, however suffer from a number of drawbacks, such as high losses \cite{CambridgeJournals8669489} and limited tunability of electronic carrier concentration. In recent years, graphene \cite{doi:10.1038/nmat1849} has emerged as a very efficient plasmonic material in the far infrared and terahertz range \cite{PhysRevB.80.245435}. Because of its unique bandstructure, electrons in graphene behave as Dirac fermions. A consequence of this is that backscattering of electrons from impurities is forbidden \cite{10.1143/JPSJ.67.1704}, which results in graphene plasmons being much less lossy in the far infrared compared to metals in the visible range. In addition to chemical doping, the carrier density or equivalently, the Fermi level in graphene can be tuned via electrostatic gating \cite{10.1126/science.1102896}. Thus graphene is an excellent candidate for tuning light-matter interaction in this wavelength range.

In this work, we discuss how decay rate can be engineered via plasmon modes in a dimer of vertically stacked graphene nanodiscs. Plasmons in a single Graphene disc have been shown to provide very high total decay rate enhancements \cite{doi:10.1021/nl201771h}. A dimer system of nanodiscs, while still having these advantages, provides a route to engineer radiative decay rates via excitation of dark and bright modes. Such bright dipolar modes have recently been experimentally observed in the case of graphene micro-disc dimers \cite{1367-2630-14-12-125001}.

We will firstly solve for the eigenmodes of the systems using a quasistatic approximation. Secondly, a general recipe using the eigenresponse theory will be provided which can be used to model the spatial and polarization dependence of the local density of states. Some of these results for the lowest dipolar mode will be compared with full-wave boundary element simulations.

Finally we use the example of the lowest dipolar mode to show that the fluorescence quantum yield can be tuned by modifying the Fermi level of the graphene nanodiscs and the possibility of obtaining vacuum Rabi splitting in the cavity.
\begin{figure}[t]
\centering
\includegraphics[width=3.5in]{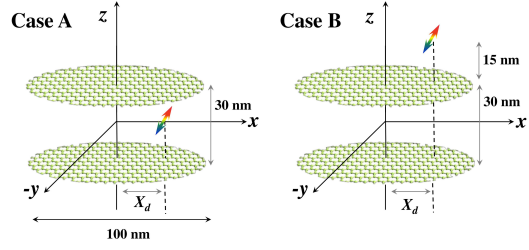}% Here is how to import EPS art
\caption{\label{fig:purcell_schematic} {\it Geometry for studying decay rate engineering:} A) In this geometry, the emitter can excite only one of the two modes, depending on its location and polarization B) In this geometry, the emitter can excite both the dark as well as the bright modes. Hence it is suited to studying comparatively, the effect of these modes on the the decay rate of the quantum emitter. The numerical values of the disc separation $D$ and the radius $R$, which were used for the BEM simulation are shown here.}
\end{figure}

\section{Methods}
In this section, we briefly mention the various analytical and numerical techniques that were employed to arrive at the results in this paper.
\subsection{Modeling electrodynamic response of graphene}
The graphene response is modeled using local random phase approximation (Local RPA). At a temperature T, the two dimensional conductivity of graphene is given by \cite{10.1140/epjb/e2007-00142-3}:
\begin{flalign}
\sigma_{RPA}(\omega) =& \frac{2e^2 kT}{\pi\hbar^2}\frac{i}{\omega+i/\tau}\ln\left|2\cosh\left(\frac{\mu}{2kT}\right)\right| \nonumber \\
&+ \frac{e^2}{4\hbar}\left[H(\omega/2, T) + \frac{4i\omega}{\pi}\int_0^{\infty} d\zeta \frac{H(\zeta,T) - H(\omega/2, T)}{\omega^2-4\zeta^2} \right]
\label{eq:RPA}
\end{flalign}
where $H(\omega,T)=\sinh(\hbar\omega/kT)/[\cosh(\mu/kT)+\cosh(\hbar\omega/kT)]$.

The first term in Eq. (\ref{eq:RPA}) represents intraband contribution and the remaining terms are contributions of the interband transitions to the total graphene conductivity. Here $\tau$ is the electron relaxation time. The relaxation time typically has contributions from 1) scattering from impurities in infinite graphene, 2) coupling to acoustic and optical phonons ($\hbar\omega_{OPh}= 0.2$ eV) in graphene and phonon modes of polar substrates and 3) edge scattering in the case of finite nanostructures, such as the one discussed in the present paper(  \cite{10.1109/JPROC.2013.2260115} and references therein). In literature, relaxation times as high as 1000 fs have been reported  \cite{10.1038/nnano.2010.172, 10.1016/j.ssc.2008.02.024}. For frequencies larger than the optical phonon frequency of graphene, $\tau\sim 50$ fs  \cite{PhysRevB.80.245435}. For the results of this paper, we use the conservative lower end of this range. For all the results concerning excitation of the disc modes via local emitters, we use $\tau = 50$ fs and $T = 300$ K. For the plane-wave excitation result, a larger $\tau$ of $100$ fs is used since for smaller values, the extinction peak for the dark mode is too broad to be separated out from the background, dominated by the bright mode. However, both these values we used for $\tau$ are on the very conservative side of the range of experimentally measured values.

\subsection{Simulation of graphene plasmon modes}
All the simulations of this paper were performed using a Boundary Element Method (BEM) code, {\sc scuff-em} suite \cite{SCUFF-EM}, a free, open-source software implementation of the boundary-element method that implements specialized algorithms for efficient computation of scattered and absorbed power in scattering problems \cite{SCUFF-PAPER}. For simulations we consider a very small thickness ``effective graphene" \cite{Graphene_Transf_Optics}. This now becomes a 3D structure, whose conductivity is given by dividing the 2D conductivity by the thickness of the “effective graphene”. This allows us to define a permittivity for “effective graphene” using Maxwell’s Equations:
\begin{equation}
\epsilon(\omega) = 1+\frac{{\it i}\sigma^{2D}(\omega)}{\omega\epsilon_{0}\Delta}
\label{eq:3D_permittivity}.
\end{equation}

where $\Delta$ is the thickness of the effective graphene. Convergence tests were performed with $\Delta$ as a parameter and a value of $\Delta= 0.25$ nm was chosen as the appropriate thickness for the specific range of frequencies and lengthscales of our problem.

\subsection{Calculation of decay rates}
In the second half of the paper, we will discuss the decay rates of emitters placed close to or inside the nanodisc cavity. An ideal dipole emitter can get rid of its energy through two pathways: 1) radiatively into free space propagating modes or 2) non-radiatively into material absorption. The decay rate into the plasmon mode is mostly dominated by absorption. We calculate these decay rates as follows. The total decay rate is calculated using the scattered electric field at the location of the emitter \cite{NovotnyBook}:
\begin{equation}
\frac{\Gamma_{tot}}{\Gamma_0} = 1 + \frac{6\pi\epsilon_0}{|\hat{{\mathbf{\mu}}}|^2 k^3}\Im\{\hat{{\bf \mu}}^{*}\cdot {\bf E_s}({\bf x_0})\}
\label{eq:GammaTot}.
\end{equation}
where $\Gamma_0$ is the decay rate of the emitter, if it were in free space.

The radiative decay rate is given by $\Gamma_{rad}/\Gamma_0 = P_{rad}/P_0 = 1 + P_{sca}/P_0$, where $P_{rad}$ is the power radiated to the far field, $P_{sca}$ is the total scattered power and $P_0$ is the power radiated by the emitter when placed in free-space. The non-radiative decay rate, which is the dominant contribution from the decay into the plasmon mode is given by $\Gamma_{abs}/\Gamma_0 = P_{abs}/P_0 \approx \Gamma_{plasmon}/\Gamma_0$, where $P_{abs}$ is the power absorbed in the graphene nanostructure.

\underline{Vacuum Rabi splitting calculation}: The most common way for describing atom-cavity interaction quantum mechanically is through the Jaynes-Cummings Model (JCM) Hamiltonian. 
The JCM Hamiltonian, in the rotating wave approximation (RWA) is given by:
\begin{equation}
H = \hbar\omega_0\sigma^{+}\sigma^{-}+\hbar\omega\left(a^{\dagger}a+\frac{1}{2}\right)+\frac{\hbar g}{2}(a^{\dagger}\sigma+a\sigma^{+})
\end{equation}
where $\omega_0$ is resonant frequency of the quantum emitter and $\omega$ is the frequency of the plasmon mode. $\sigma^{+}$ ($\sigma^{-}$) are the atomic raising (lowering) operators and $a^\dagger$ ($a$) are the creation (annihilation) operators for a cavity photon.

For the present problem we use an open quantum system approach, in order to incorporate absorption and radiative decay. Thus the evolution of the density matrix is given by \cite{10.1103/PhysRevB.87.115419}:
\begin{flalign}
\frac{d\rho }{dt}= &-\frac{i}{\hbar}[H,\rho]-\frac{\kappa}{2}(a^{\dagger}a\rho - 2a\rho a^{\dagger} +\rho a^{\dagger} a)\nonumber \\
&- \frac{\Gamma^{'}}{2}(\sigma^{+}\sigma^{-}\rho - 2\sigma^{-}\rho \sigma^{+}+\rho \sigma^{+}\sigma^{-})
\label{eq:solve}
\end{flalign}
where $\kappa$ is the rate of decay of the plasmon mode. $\kappa$ contains both the radiation as well and absorption mechanisms for broadening the plasmon resonance \cite{10.1103/PhysRevA.82.043845}. However, it is usually dominated by absorption. $\Gamma'$ is the decay rate of the quantum emitter into free space modes, modified by geometrical effects.  However, in this paper we assume the $\Gamma'$ to be equal to $\Gamma_0$, the spontaneous emission rate in free space, in accordance with Wigner-Weisskopf theory. 

The system density operator evolves according to Eq. (\ref{eq:solve}). In the single excitation manifold, only the  states \{$\Ket{g}\bigotimes\Ket{1}$, $\Ket{e}\bigotimes\Ket{0}$, $\Ket{g}\bigotimes\Ket{0}$\} need to be retained. Here $\Ket{g}$ and $\Ket{e}$ are the ground and excited states of the atom and $\Ket{0}$ and $\Ket{1}$ denote the number of photons in the cavity mode. It can then be shown \cite{Solid-State-QED} that for Rabi oscillations to exist, on resonance ($\Delta=0$), one needs the condition $|g/(\kappa-\Gamma')| >1/2$.

In the JCM, the coupling strength $g$ is determined by the details of the cavity field mode and the atomic dipole matrix element. For our purpose, we determine $g$ classically, using the limit of a low finesse cavity. In this limit, $g$ satisfies the following equation:
\begin{equation}
\Gamma_{tot} = \Gamma_0+\frac{g^2 (\kappa-\Gamma')}{4\Delta^2+(\kappa-\Gamma')^2}
\label{eq:bad_cavity}
\end{equation}
where $\Delta$ is the detuning between the resonant frequency of the plasmon mode and that of the quantum emitter.
For the present work, the typical spontaneous emission rate of the emitter is much smaller than the cavity line-width. Thus, Eq. (\ref{eq:bad_cavity}) suggests that on resonance, $\Gamma_{tot}/\kappa = (g/\kappa)^2$. Hence the $g$ factor can be determined. This expression also points out that Rabi oscillations should exist when $\Gamma_{tot}/\kappa > 1/4$.
\section{Results and discussion}
\subsection{Calculation of the eigen-modes in the quasistatic limit}
Firstly, we discuss the mathematical formulation of the eigenvalue equation in terms of an electrostatic potential. This is essentially a solution of the Laplace equation for the disc geometry. This section is divided into two subsections. In the first subsection, we repeat the derivation for the case of single disc, which had been worked out by Fetter\cite{PhysRevB.33.5221} in 1986. In the second subsection, we formulate the eigenvalue problem for the case of a stack of two discs.

In the second section, we discuss a numerical framework to solve the eigenvalue equations we obtained in the previous section. In the third section, we summarize the results of the calculation, by providing details of the eigen-mode plots and a comparison to full wave boundary element simulations.

\subsubsection{Mathematical framework for a stacked dimer of discs}
The general strategy is to solve the Poisson equation for the electrostatic potential $\Phi(\mathbf{r})$, with the surface charge boundary condition due to the graphene discs. The surface charge density itself can be related to the electrostatic potential via the continuity equation and the surface conductivity of graphene. This leads to an eigenvalue equation with $\Phi$ on both sides. Subsequently, numerical techniques are used to solve this eigenvalue problem to get the resonant frequencies as well as the potential profile. This potential can then be used to calculate various other quantities of interest such as surface polarization and surface current density.

The calculation for the single disc case, was carried out in the Fetter's paper\cite{PhysRevB.33.5221} on magnetoplasmons in disk geometries of 2DEG. For the sake of comparison the notation from a recent paper\cite{PhysRevB.86.125450} on edge plasmons in a single graphene disc, has been used here. 

It should also be noted that for a single disc in the quasistatic regime, closed form solution is possible\cite{:/content/aip/journal/jcp/77/12/10.1063/1.443833}. However, we use a numerical approach here which can be easily extended to stacks consisting of arbitrary number of discs, where closed form solution becomes very cumbersome.

The geometry consists of two identical discs each of radius $R$, stacked vertically with a separation $D$ in between (see Fig. \ref{fig:purcell_schematic}). The location of the discs in our chosen coordinate system is $z=\pm D/2$. The approach that will be presented here can easily incorporate the case where the two discs are non-identical. However, for the sake of clarity for our specific case, we will only consider identical discs for now.

First note that because of circular symmetry, the potential can be expressed as $\Phi(\mathbf{r}) = \Phi(r,z)e^{iL\phi}$, in cylindrical coordinates.

We follow a two step procedure to get to the eigenvalue equation:
\begin{itemize}
\item Express the surface potential $\Phi(r,z= \pm D/2)$ in terms of the surface charge density $\sigma_b$, using the Laplace equation and the normal electric field boundary condition
\item Express surface charge density $\sigma_b$ in terms of the surface potential $\Phi(r, z= \pm D/2)$, using the continuity equation and the current-field relation
\end{itemize}

\underline{Expressing $\Phi$ in terms of $\sigma_b$}: 
The Poisson equation in this case is given by:
\begin{equation}
\nabla^2\Phi(r,z) = -\frac{\sigma_b\Theta(R-r)}{\epsilon_0}(\delta(z-D/2) + \delta(z+D/2))
\end{equation}
where $\sigma_b$ is the surface charge density (and not the surface conductivity, which is represented by $\sigma$).

One way of solving such problems is to write the general form of the solution in the regions on either side of the the boundaries and then match the boundary conditions. We will use this approach.

Thus we write the solution for the Laplace equation for $z \neq 0$ and then use the boundary condition for the normal electric field. To be specific, these equations are given below:
\begin{equation}
\nabla^2\left[\Phi(r,z)e^{\imath L\phi}\right] = 0
\label{eq:laplace_eqn}
\end{equation}
\begin{equation}
\epsilon_u\frac{\partial\Phi(r, z)}{\partial z}\biggr |_{(\frac{D}{2})^+} - \epsilon_m\frac{\partial\Phi(r, z)}{\partial z}\biggr |_{(\frac{D}{2})^-} = -\frac{\sigma_{b,u}\Theta(R-r)}{\epsilon_0}
\end{equation}
\begin{equation}
\epsilon_m\frac{\partial\Phi(r, z)}{\partial z}\biggr |_{(-\frac{D}{2})^+} - \epsilon_d\frac{\partial\Phi(r, z)}{\partial z}\biggr |_{(-\frac{D}{2})^-} = -\frac{\sigma_{b,l}\Theta(R-r)}{\epsilon_0}
\end{equation}
where $\epsilon_m$ is the relative permittivity of the medium in between the discs. Note that the $e^{iL\phi}$ dependence was suppressed in the boundary condition equation. (Note that there is another boundary condition which is the continuity of the potential across $z=\pm D/2$.)

Now let us express $\Phi$ in terms of its Hankel transform component:
\begin{equation}
\Phi(r,z)e^{\imath L\phi} = \int_0^{\infty}dp\ p \bar{\Phi}(p,z)J_L(pr)e^{\imath L\phi}
\label{eq:forward_hankel}
\end{equation}
where the Hankel transform is only taken in the radial coordinate of the cylindrical system.

Now we substitute Eq. (\ref{eq:forward_hankel}) into Eq. (\ref{eq:laplace_eqn}). After some manipulation and using Bessel's differential equation, we obtain the following simplified form:
\begin{equation}
\int_0^{\infty}dp\ p \left[\left(\frac{\partial^2}{\partial z^2}-p^2\right)\bar{\Phi}(p,z)\right]J_L(pr)e^{\imath L\phi} = 0
\label{eq:single_disc_hankel_laplacian}
\end{equation}

For $z\neq \pm D/2$, Eq. (\ref{eq:laplace_eqn}) holds for the potential $\Phi(\mathbf{r})$ in real coordinates. Equivalently, for $z\neq \pm D/2$, Eq. (\ref{eq:single_disc_hankel_laplacian}) holds for the Hankel transformed potential. We can write down the form of the solution in the three different regions as follows:
\begin{align}
    \bar{\Phi}(p,z)= 
\begin{cases}
    A_u\  e^{-p(z-D/2)},& \text{if } z\geq D/2\\
    A_m^{+}\  e^{p(z-D/2)}+A_m^-\  e^{-p(z+D/2)},& \text{if } |z|\leq D/2\\
    A_d\  e^{p(z+D/2)},& \text{if } z\leq D/2
\end{cases}
\end{align}
There are four unknowns $A_u,\ A_m^+,\ A_m^-,\ A_d$. We also have four equations, two for the continuity of the potential across the discs and the other two for the normal electric field boundary condition.

It is quite straightforward to solve for the general case of different relative permittivities. In the following, we choose $\epsilon_u=\epsilon_m=\epsilon_d=\epsilon$ for simplicity. Solving the above linear system of equations, we get the solution for the Hankel-transformed potential on the discs:
\begin{equation}
\renewcommand{\arraystretch}{1.5}
\begin{bmatrix}
\bar{\Phi}_u(p)\\
\bar{\Phi}_d(p)
\end{bmatrix}=\frac{1}{2p\ \epsilon_0\epsilon}
\renewcommand{\arraystretch}{1.5}
\begin{bmatrix}
1 & e^{-pD} \\
e^{-pD} & 1
\end{bmatrix}
\renewcommand{\arraystretch}{1.5}
\begin{bmatrix}
\bar{\sigma}_{b,u}(p)\\
\bar{\sigma}_{b,d}(p)
\end{bmatrix}
\end{equation}
Now we go to real space, by taking the inverse Hankel transform on each side of the above equation. For brevity, we denote the Hankel transform operator as $\hat{\mathfrak{H}}(p;r') = \int_0^{\infty}dr\ r' J_L(pr')$ and its inverse operator as $\hat{\mathfrak{H}}^{-1}(r;p) = \int_0^{\infty}dp\ p J_L(pr)$. With this notation, we can express the real space solution as:
\begin{equation}
\renewcommand{\arraystretch}{1.5}
\begin{bmatrix}
\Phi_u(r)\\
\Phi_d(r)
\end{bmatrix}=\frac{1}{\epsilon_0\epsilon}
\renewcommand{\arraystretch}{1.5}
\begin{bmatrix}
\int_0^{R}dr'\ r' K_L^{o}(r,r') & \int_0^{R}dr'\ r' K_L^i(r,r') \\
\int_0^{R}dr'\ r' K_L^i(r,r') & \int_0^{R}dr'\ r' K_L^o(r,r')
\end{bmatrix}
\renewcommand{\arraystretch}{1.5}
\begin{bmatrix}
{\sigma}_{b,u}(r')\\
{\sigma}_{b,d}(r')
\end{bmatrix}
\label{eq:double_disc_step_1}
\end{equation}
with $K_L^o(r,r')=\frac{1}{2}\int_0^{\infty}dp J_L(pr)J_L(pr')$ being the kernel for on-site term and $K_L^i(r,r')=\frac{1}{2}\int_0^{\infty}dp\ e^{-pD} J_L(pr)J_L(pr')$ for the interaction term.

\underline{Expressing $\sigma_b$ in terms of $\Phi$}: 
There are two equations that we need to express $\sigma_b$ in terms of $\Phi$. One is the continuity equation for surface current density and the other is the relation between surface current density and the electric field. These equations are given below:
\begin{equation}
\nabla_{||}\cdot\mathbf{J}_s + \frac{\partial\sigma_b}{\partial t}=0
\end{equation}
\begin{equation}
\mathbf{J}_s = \sigma(\omega)\mathbf{E}_{||}
\end{equation}
Now using the relation $\mathbf{E}_{||} = -\nabla_{||}(\Phi(r, z=0)e^{\imath L \phi})$, we arrive at the relation:
\begin{equation}
\renewcommand{\arraystretch}{1.5}
\begin{bmatrix}
{\sigma}_{b,u}(r')\\
{\sigma}_{b,d}(r')
\end{bmatrix}=-\frac{\sigma(\omega)}{\imath\omega}
\renewcommand{\arraystretch}{1.5}
\begin{bmatrix}
\Theta(R-r'){\nabla'}_{||}^2-\delta(r'-R)\frac{\partial}{\partial r'} & 0 \\
0 & \Theta(R-r'){\nabla'}_{||}^2-\delta(r'-R)\frac{\partial}{\partial r'}
\end{bmatrix}
\renewcommand{\arraystretch}{1.5}
\begin{bmatrix}
{\Phi}_u(r')\\
{\Phi}_d(r')
\end{bmatrix}
\label{eq:double_disc_step_2}
\end{equation}
Thus, combining Eq. (\ref{eq:double_disc_step_1}) and Eq. (\ref{eq:double_disc_step_2}), we arrive at the final eigenvalue equation for the stacked discs case:
\begin{flalign}
\renewcommand{\arraystretch}{1.5}
\begin{bmatrix}
\Phi_u(r)\\
\Phi_d(r)
\end{bmatrix}=\frac{\sigma(\omega)}{\imath\omega\epsilon_0\epsilon}
\renewcommand{\arraystretch}{1.5}
&\begin{bmatrix}
\int_0^{R}dr'\ r' K_L^{o}(r,r') & \int_0^{R}dr'\ r' K_L^i(r,r') \\
\int_0^{R}dr'\ r' K_L^i(r,r') & \int_0^{R}dr'\ r' K_L^o(r,r')
\end{bmatrix}\cdot\nonumber \\
\renewcommand{\arraystretch}{1.5}
&\begin{bmatrix}
-\Theta(R-r'){\nabla'}_{||}^2+\delta(r'-R)\frac{\partial}{\partial r'} & 0 \\
0 & -\Theta(R-r'){\nabla'}_{||}^2+\delta(r'-R)\frac{\partial}{\partial r'}
\end{bmatrix}%\nonumber \\
\renewcommand{\arraystretch}{1.5}
\begin{bmatrix}
{\Phi}_u(r')\\
{\Phi}_d(r')
\end{bmatrix}
\end{flalign}

As the last step, we move to normalized coordinates $x\rightarrow r/R$ and express the final equation in operator form:
\begin{equation}
\renewcommand{\arraystretch}{1.5}
\begin{bmatrix}
\phi_u(x)\\
\phi_d(x)
\end{bmatrix}
=\eta
\renewcommand{\arraystretch}{1.5}
\begin{bmatrix}
\hat{\mathcal{I}}^{uu}_L(x;x') & \hat{\mathcal{I}}^{ud}_L(x;x')\\
\hat{\mathcal{I}}^{du}_L(x;x') & \hat{\mathcal{I}}^{dd}_L(x;x')\\
\end{bmatrix}
\renewcommand{\arraystretch}{1.5}
\begin{bmatrix}
\phi_u(x')\\
\phi_d(x')
\end{bmatrix}
\label{eq:DoubleDiscFredholm}
\end{equation}
where $\eta = \sigma(\omega)/\imath\omega\epsilon_0\epsilon R$. The reader is reminded that here $\hat{\mathcal{I}}^{uu}_L(x;x')$ and $\hat{\mathcal{I}}^{dd}_L(x;x')$ are associated with the on-site term for the upper and lower disc respectively, whereas the off-diagonal terms represent the interaction between the two discs. In terms of normalized coordinates, $\Phi(r, z= \pm  D/2)\rightarrow\phi_{u,d}(x)e^{\imath L\phi}$.
Eq. (\ref{eq:DoubleDiscFredholm}) is an eigenvalue problem in the parameter $\eta=\sigma(\omega)/(i\omega\epsilon_0\epsilon R)$, which can be related to the resonant frequencies of the modes.
 Note that in the normalized coordinates, the exponential term in the off-diagonal kernel depends on the ratio $D/R$ instead of just $D$.
 
The solution to Eq. (\ref{eq:DoubleDiscFredholm}), will give us the resonant frequencies and the mode-profiles of the plasmons in stacked dimer of graphene discs.

It should be noted that this kind of approach is easily extensible to more than two discs or discs with different radii or surface conductivities.

We solve the eigenvalue problems for the single disc and the stacked dimer of discs case using the standard method of polynomial expansion. The results for the eigen-frequencies are shown in Fig. \ref{fig:ResFreqTogether}.

\subsubsection{Comparison with full wave simulation}
\begin{figure}
\centering
  \includegraphics[width=5.0in]{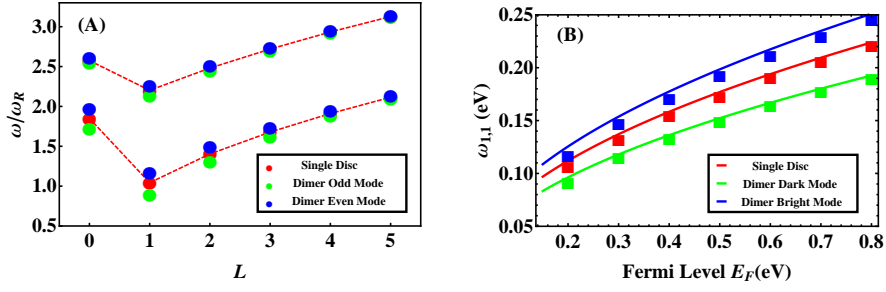}
\caption{\label{fig:ResFreqTogether} {\it A) Resonant frequencies of the modes calculated using the quasistatic solution } Normalized resonant frequencies of the modes of different symmetry, where $\omega^2_R = e^2 E_F / (2\pi\hbar^2\epsilon_0\epsilon R)$. For each $L$, only the lowest two modes are shown. The dashed line is only a guide to the eye. {\it B) Comparison of the resonant frequencies calculated using quasistatic approximation versus the full wave boundary element simulation } Resonant frequencies of the $L=1,\ n=1$ modes as a function of Fermi level $E_F$. Squares represent the BEM calculation and lines represent the quasistatic result. Note that in BEM we chose finite absorption whereas in the quasistatic approach we used a lossless graphene conductivity.}
\end{figure}

We compared the resonant frequencies obtained using the quasi-static solution to those obtained using a full-wave boundary element simulation (BEM). For this comparison we only choose the $L=1,\ n=1$ mode since that is the mode that we will be concerned with in the rest of the paper, when talking about photon emission rate engineering. It should also be noted that for the simulations, we use a realistic absorption in the graphene conductivity. The comparison is presented in Fig.\ref{fig:ResFreqTogether}.
Figure \ref{fig:ResFreqTogether} suggests that there is a good overall match between the resonant frequencies found from the BEM and the quasistatic result. The resonant frequencies in the simulation were obtained from the LDOS spectrum. 

Two features in Fig. \ref{fig:ResFreqTogether} are worth highlighting. Firstly, the resonant frequencies of both the modes increase with $E_F$. This is due to the fact that increasing $E_F$ results in an increase in the carrier density, which in turn causes an increase in the restoring force. This explanation is similar to how the plasma frequency in noble metals increases with carrier concentration. Secondly, the frequency splitting $\omega_B-\omega_D$ increases with $E_F$. This is due to the fact that at higher $E_F$, the plasmon modes of individual discs are more leaky. This results in the interaction between the two discs being even stronger, resulting in a larger splitting.

\subsection{Eigen-response theoretic framework and calculation of overlap}
To understand dependence of the spatial and polarization dependence of the LDOS, we resort to an eigen-response theory \cite{PhysRevB.89.045408}. In the present case of a system with a symmetric and an antisymmetric mode, the polarization density can be expressed as:
\begin{equation}
\Ket{P} = \sum_{L=0}^{\infty}\alpha_{A,L}\Ket{P_{A,L}}\Braket{P_{A,L}|E_{exc}} + \alpha_{S,L}\Ket{P_{S,L}}\Braket{P_{S,L}|E_{exc}}
\end{equation}
where $\alpha_{A,L}$ and $\alpha_{S,L}$ are the eigen-polarizabilities of the antisymmetric and symmetric modes and $P_{A,L}$ and $P_{S,L}$ are the eigenmodes.

The excitability of the modes is related to the overlap terms, which are the inner products of the mode profile and the excitation. 

\underline{Relation between overlap and LDOS}: 
The spatial dependence of the LDOS is contained in the overlap term, since the eigen-polarizability is usually only frequency dependent. In general the projected LDOS in Eq. (\ref{eq:GammaTot}) can be written as:
\begin{equation}
\Gamma_{tot}/\Gamma_0 =1 +6\pi k \sum_{i}\left | \Braket{\mathbf{\mu}(\mathbf{r}_0)|\hat{\mathbf{G}}(\mathbf{r}, \mathbf{r}_0) | i} \right |^2 \Im[\alpha_{i}(\omega)]
\end{equation}
where $\Ket{i}$ is a quantity proportional to the surface polarization, for each mode. In the following section we present the calculation of the surface polarization, which will help us calculate these overlap terms.

\underline{Calculation of surface polarization}: 
From our quasistatic approach, we determined the surface potential on both the discs. Using the surface potential, it is straightforward to obtain the surface polarization.
The surface polarization $\mathbf{P}_s$ can be related to the potential on the discs in the following way. In the absence of magnetization, the surface current $\mathbf{J}_s$ can be related to $\mathbf{P}_s$ as follows:
\begin{equation}
\mathbf{J}_s = \frac{\partial\mathbf{P}_s}{\partial t} = -\imath\omega\mathbf{P}_s
\end{equation}
$\mathbf{J}_s$ can also related to the electrostatic potential using the relation:
\begin{equation}
\mathbf{J}_s=[\sigma(\omega)\Theta(R-r)]\mathbf{E}_{||} = -[\sigma(\omega)\Theta(R-r)]\nabla_{||}\Phi
\end{equation}
Thus we have the relation between $\mathbf{P}_s$ and $\Phi$:
\begin{equation}
\mathbf{P}_s = \left[\frac{\sigma(\omega)\Theta(R-r)}{\imath\omega}\right]\nabla_{||}\Phi
\label{eq:Ps_Phi}
\end{equation}
The potential $\Phi$ obtained by solving the eigenvalue equation as mentioned in an earlier note, can be plugged in Eq. (\ref{eq:Ps_Phi}) to calculate $\mathbf{P}_s$ and subsequently the overlap terms.

We will consider the source dipole generating a field which excite various infinitesimal dipoles on the disc surface. For this purpose, we will need the Green's tensor which is defined as\cite{NovotnyBook}:
\begin{equation}
\hat{\mathbf{G}}(\mathbf{r},\mathbf{r}_0) = \frac{k}{4\pi} \left(A(k|\mathbf{r}-\mathbf{r}_0|) + B(k|\mathbf{r}-\mathbf{r}_0|) \frac{|\mathbf{r}-\mathbf{r}_0\rangle \langle\mathbf{r}-\mathbf{r}_0 |}{|\mathbf{r}-\mathbf{r}_0|^2} \right)
\end{equation}
In order to keep the mathematical framework completely general, we consider the geometry shown in Fig. \ref{fig:purcell_schematic}.  For this geometry, $\mathbf{r}_0 = X_d \hat{x} + Y_d \hat{y} + Z_d \hat{z}$ and the location of the infinitesimal dipoles $\mathbf{r} = x\hat{x} + y\hat{y} + z \hat{z}$. 

\begin{figure}
\centering
  \includegraphics[width=5.0in]{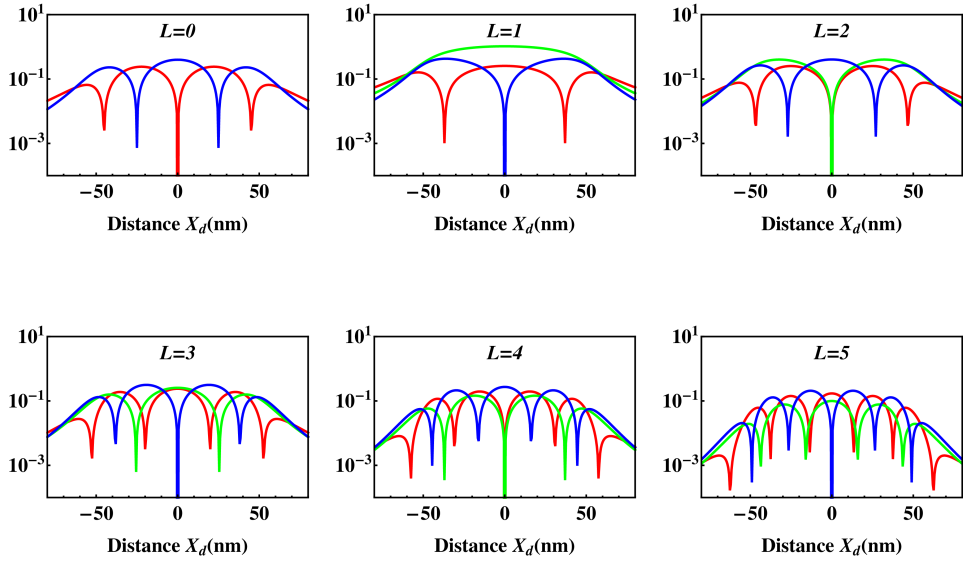}
\caption{\label{fig:AllSingleDiscPlots} {\it Semi-analytically calculated overlap terms for a Single Disc:} The emitter is located at a vertical distance of $15$nm above the disc and moves along $Y_d=0$.  The colors correspond to different polarizations of the emitter: $\hat{x}$(\textcolor{red}{{\Large{{{\textbf{\textrm{--}}}}}}}), $\hat{y}$(\textcolor{green}{{\Large{{{\textbf{\textrm{--}}}}}}}) and $\hat{z}$(\textcolor{blue}{{\Large{{{\textbf{\textrm{--}}}}}}})}
\end{figure}

The overlap terms for the case of a single disc are shown in Fig. \ref{fig:AllSingleDiscPlots}.
\subsection{Decay rate engineering}
Single graphene nanodiscs, have been shown to provide very high Purcell factors \cite{doi:10.1021/nl201771h}. Such high enhancement factors are possible due to the very small plasmon mode volume, which is a general characteristic of graphene films and nanostructures in the far infrared and terahertz range \cite{10.1038/nphoton.2012.262}. Using a vertical nanodisc dimer cavity, should provide an additional degree of freedom for engineering the decay rates. For instance, other than applying gate voltage or changing the radius of the discs, there is now an additional parameter, which is the separation of the discs, that can be used to tune the resonances \cite{10.1103/PhysRevB.81.085444}.

The dark dipolar plasmon mode only weakly couples to plane waves hence its excitation using plane wave is not very efficient. However, if we use local emitters, such as quantum dots, it is indeed possible to excite the dark mode very efficiently \cite{10.1103/PhysRevB.81.085444}. The coupling of quantum emitters to such plasmon modes is reflected in the modification of the decay rates of the former. Depending on the nature of the plasmon mode, radiative decay of an emitter can be suppressed (quenching) or enhanced. In the case of graphene nanodisc dimer cavities, one can easily modify the radiative or non-radiative processes, by using gate voltage or disc separation to tune the dark or the bright mode to be resonant to the quantum dot. In this context, location and polarization of the emitter is another variable \cite{10.1364/OL.37.001017}, which will be discussed in this work.

We consider two situations (Fig. \ref{fig:purcell_schematic}) for studying the coupling between the emitter and the plasmon modes. Firstly, we study the case when the emitter placed inside the cavity. If the emitters are located at the inversion plane then it allows us to excite either the dark or bright plasmon mode, depending on the emitter polarization and position. We will use the eigen-response theory to calculate the overlap terms as a function of the position and polarization of the emitter. We will also present a comparison with LDOS for the lowest dipolar mode calculated via BEM simulation and show that the shape of the LDOS spectrum can be well explained by the calculated overlap terms. Secondly, we place the emitter outside the cavity, resulting in it being able to excite both the modes, for the same location and orientation of the emitter's dipole moment vector. This enables us to directly compare the bright and dark modes, in terms of their efficiency in suppressing or enhancing the decay pathways and quantum efficiency of the emitter, under similar excitation conditions. To avoid repetition, in this section we will only present simulation results for the radiative decay rates.

\subsubsection{Case A: Emitters inside the stacked disc cavity}
\begin{figure}[htb]
\centering
  \includegraphics[width=5.0in]{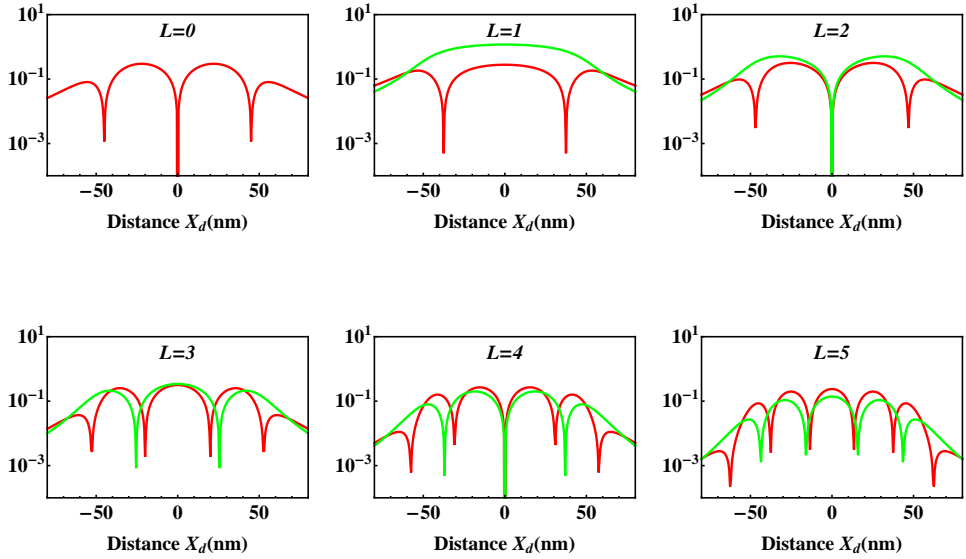}
\caption{\label{fig:AllDoubleDiscBrightPlots}{\it Semi-analytically calculated overlap terms for the disc dimer bright mode:} The emitter is located on the inversion plane parallel to the discs and moves along $Y_d=0$. The colors correspond to different polarizations of the emitter: $\hat{x}$(\textcolor{red}{{\Large{{{\textbf{\textrm{--}}}}}}}) and $\hat{y}$(\textcolor{green}{{\Large{{{\textbf{\textrm{--}}}}}}}). The $\hat{z}$-polarization has a zero overlap at these resonances.}
\end{figure}
In principle, one can calculate the decay rates for all emitter positions inside the cavity. A randomly oriented emitter should then couple to both the dark as well as a bright mode. However, if the quantum dots are placed inside the cavity, on the inversion symmetry plane parallel to the plane of the disks, then depending of the alignment of the dipole matrix element, the quantum dot will only be able to couple to either the dark or bright mode, but not both. Hence in order to understand the physics of the problem, this is a convenient choice of emitter location.

\underline{Position and polarization dependence of the LDOS}:
Having worked out the matrix elements for the case of the single disc, we now move on to consider the cases of the dimer of discs. Because of mirror symmetry we can predict that there are even and odd modes. In literature, these are often called the bright and dark modes respectively. If we label the upper disc by U and the lower disc by L, then the following relations hold:
\begin{itemize}
\item Bright mode: $d\mathbf{p}_U = d\mathbf{p}_L$, $z_{U} =  -z_{L} = d/2$
\item Dark mode: $d\mathbf{p}_U = - d\mathbf{p}_L$, $z_{U} = - z_{L} = d/2$
\end{itemize}
Now to calculate the total overlap for the two modes, we add the contributions from the two discs, taking into account the appropriate sign changes as mentioned in the relations given above:\\
{\it Bright Mode}:
Since the sign of the infinitesimal dipole moment does not change, we only need to consider the sign change in the $z$ coordinate of the two discs. This results in the total sum of the overlap term for the $z=\pm d/2$ giving a zero for the $z-$polarization of the emitter. The other two terms for the $x$ and $y$ polarizations survive and are basically twice of the corresponding value for the single disc case.\\
{\it Dark Mode}:
In this case, the sign of the infinitesimal dipole moment does change, and so does the sign change in the $z$ coordinate of the two discs. This results in the total sum of the overlap term for the $z=\pm d/2$ giving a zero for the $x-$ and $y-$polarizations of the emitter. The only nonzero term is the one for the $z-$polarizations and is just twice of the corresponding value for the single disc case.

\begin{figure}
\centering
  \includegraphics[width=5.0in]{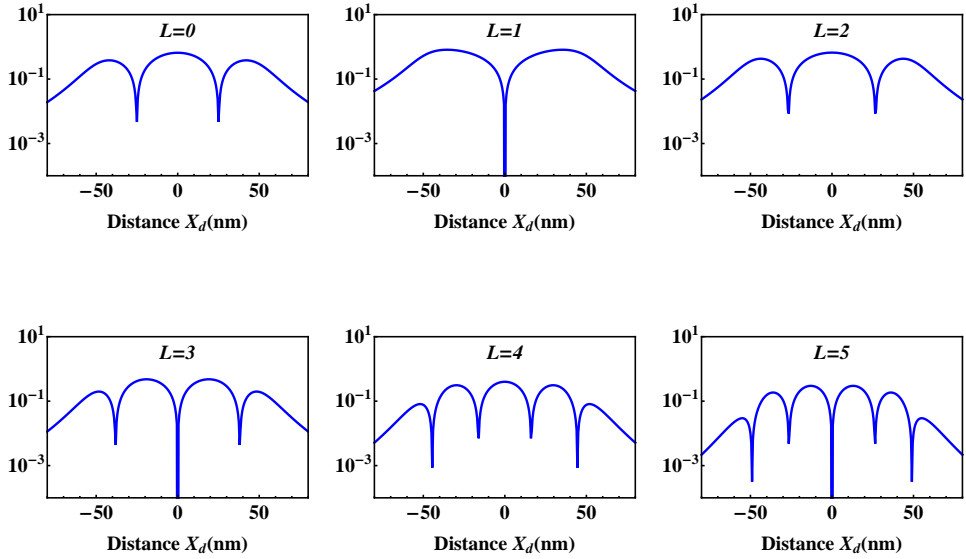}
\caption{\label{fig:AllDoubleDiscDarkPlots}{\it Semi-analytically calculated overlap terms for the disc dimer dark mode:} The emitter is located on the inversion plane parallel to the discs and moves along $Y_d=0$. The color (\textcolor{blue}{{\Large{{{\textbf{\textrm{--}}}}}}}) corresponds to the $\hat{z}$-polarization of the emitter. The $\hat{x}$ and $\hat{y}$-polarizations produce zero overlap at these resonances.}
\end{figure}

The overlap terms are calculated by evaluating integrals of the form $\int{d\mathbf{p}_{mode}^*\cdot \mathbf{E}_{exc}}$, as discussed earlier. The calculated overlap terms are presented in Fig. \ref{fig:AllDoubleDiscBrightPlots} for the bright mode and Fig. \ref{fig:AllDoubleDiscDarkPlots} for the dark mode.

\begin{figure}
\centering
  \includegraphics[width=4.5in]{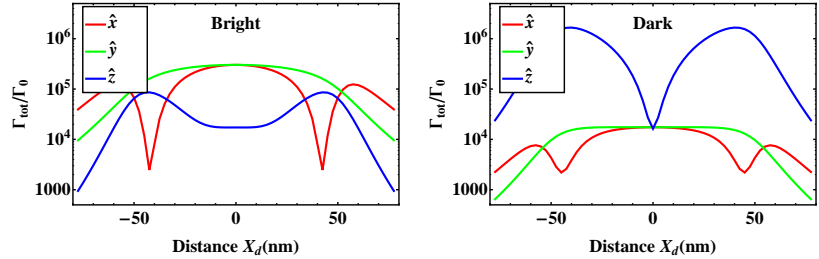}
\caption{\label{fig:bem_purcell_inside}{\it Total decay rate as a function of emitter position and polarization:} Using BEM, we calculate the total decay rate of an emitter placed at ($X_d$, 0, 0), at the bright (left) and dark (right) mode frequencies for $(L=1,\ n=1)$ mode. ($E_F=0.5$ eV and $\tau = 0.05$ ps)}
\end{figure}

To verify our approach, we perform BEM simulations for a dimer of graphene discs, each 100 nm in diameter and separated vertically by 30 nm. The frequency range for the simulation is chosen to highlight the contribution of the lowest dipolar mode.
The semi-analytical calculations using the eigen-response theory gives the dips in the LDOS spectrum( Fig. \ref{fig:AllDoubleDiscBrightPlots} for the bright mode, Fig. \ref{fig:AllDoubleDiscDarkPlots} for the dark mode ) at the correct positions, consistent with the results of the BEM simulations (Fig. \ref{fig:bem_purcell_inside}). Note that there are additional features seen in the simulation results. For instance, for the dark mode, there is an LDOS feature for the x and y-polarized emitters. Similarly for the bright mode, there is a contribution from the z-polarized emitter. This effect is due contributions from neighbouring resonances. in fact this effect can be easily taken in to account in the eigen-response theory, if we include losses in the eigen-polarizabilities. These calculations will be published elsewhere.
In the next section, we point out how radiative decay rates can be engineered using these dark and bright modes.
\subsubsection{Case B: Emitters outside the stacked disc cavity}
In order to compare quantitatively, the effect of the dark and bright modes on the quantum emitter decay rate, we now study a geometry in which the emitter can excite both modes. Any location of the emitter other than the inversion plane is a valid choice. However, for simplicity, we chose to consider the emitter located outside the cavity as shown in Fig. \ref{fig:purcell_schematic}. This might also be a convenient scheme, in as far as experimental realization is concerned. We can study all three orientations of the emitter dipole moment as done in the previous section. However, since in this section our main aim is to highlight the engineering of radiative decay rates, we will consider only one one polarization of the emitter, namely the x-polarization. To gain some qualitative insight into the excitability of the modes, we analytically calculate the overlap terms. Based on this calculation (not shown), we find that for an emitter close to the center $X_d=0$ nm, both the modes are excitable, with highest probability.

\begin{figure}
\centering
\begin{subfigure}{.5\textwidth}
  \centering
  \includegraphics[width=2.5in]{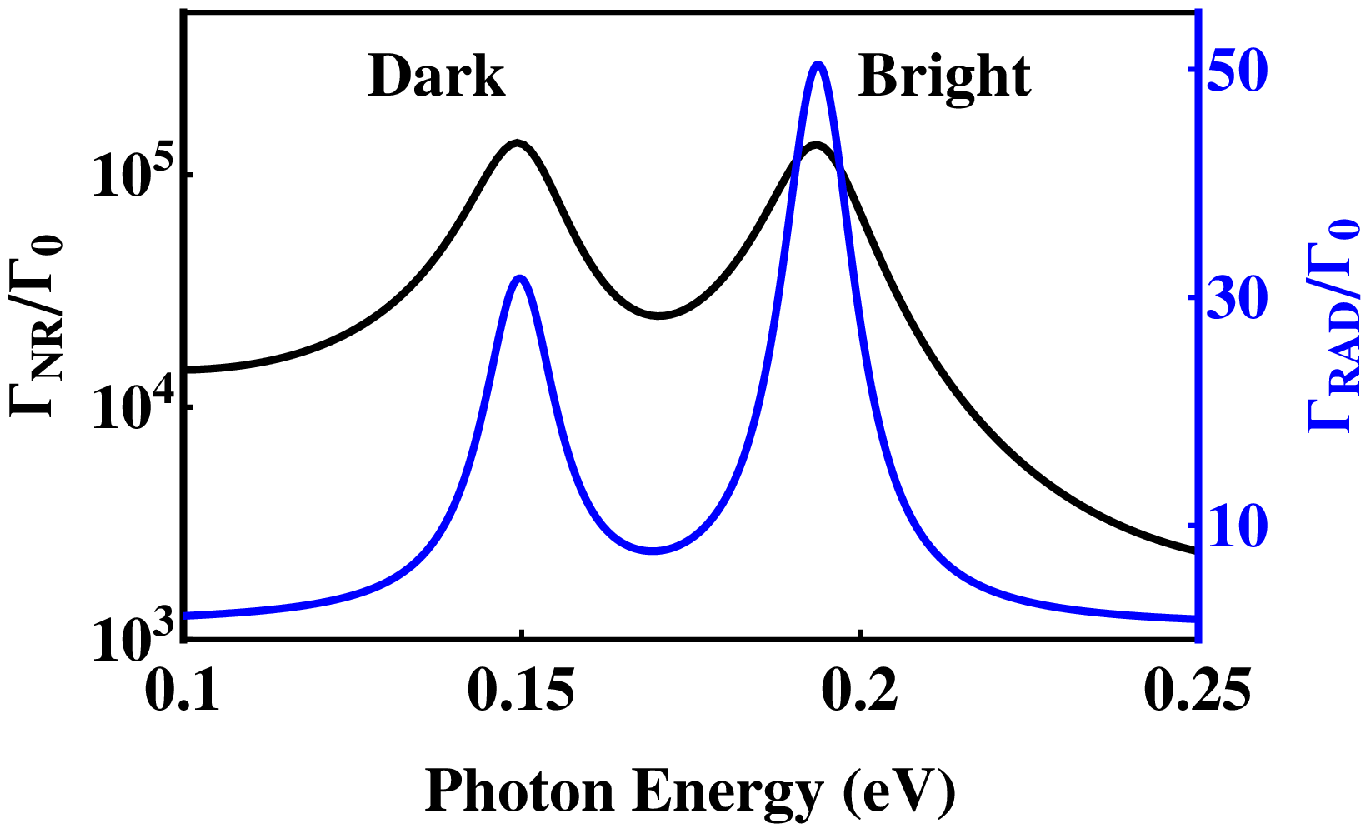}
  \caption{}\label{fig:purcell_spectrum_outside}
\end{subfigure}%
\begin{subfigure}{.5\textwidth}
  \centering
  \includegraphics[width=2.5in]{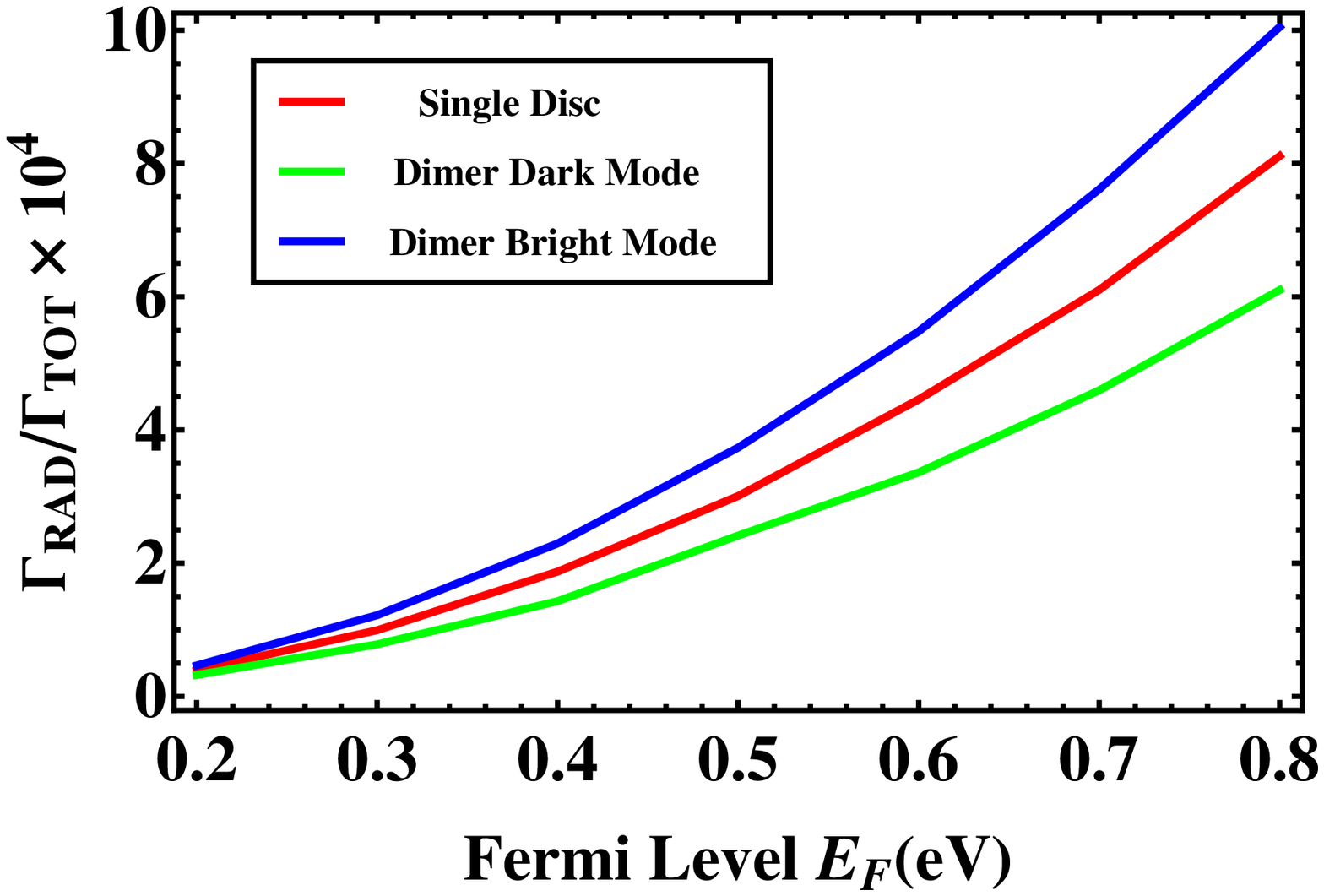}
  \caption{}\label{fig:radiative_efficiency_outside}
\end{subfigure}
\caption{\ a) {\it Simulated (BEM) Spectrum of Decay Rates } An example spectrum of the two decay rates into the radiative ($\Gamma_{RAD}/\Gamma_0$) and plasmon ($\approx\Gamma_{NR}/\Gamma_0$) modes. The dipole moment of the emitter is aligned in the x-direction and it is located at $X_d=0$ nm and $Z_d$ = 30 nm, fixed such that the emitter is located 15 nm above the closest disc. b) {\it Comparison of Radiative Efficiency: } The quantum emitter positioned at $X_d=0$ nm at a vertical distance of 15 nm above the cavity. Depending on it's resonant frequency, it can couple to both the dark as well as bright modes. It can be seen here that the dark mode suppresses radiative emission, whereas the bright mode enhances it, compared to the single disc case. ($\tau = 0.05$ ps)   }
\label{fig:emitter_outside_decay_rates}
\end{figure}

\underline{Radiative decay rate engineering}: 
We then performed simulations to calculate the radiative, non-radiative and total decay rates, when the quantum emitter, located at $(X_d, 0, Z_d)$ couples to either of the modes at their respective resonant frequencies. An example spectrum, for $X_d = 0$ nm is shown in Fig. \ref{fig:purcell_spectrum_outside}. In this case, the total decay rate enhancement, which is close to the non-radiative part, is almost the same for the two modes. This is consistent with our analytical calculation of the overlap terms which show that at $X_d = 0$ nm both the modes are equally excitable. However, there is a difference between the radiative decay rate enhancement. This situation results because of partial cancellation of the induced moments on two discs, for the dark mode. 

Further, we demonstrate the tunability of the radiative efficiency as a function of carrier concentration. Our calculations in Fig. \ref{fig:radiative_efficiency_outside} show that the contrast in radiative efficiency increases with increasing $E_F$. The dark mode becomes less, and the bright mode, more radiative, as the carrier concentration increases.  As mentioned before, this is because at higher carrier concentrations, the modes of the two discs can interact more strongly. 

\begin{figure}
\centering
\begin{subfigure}{.5\textwidth}
  \centering
  \includegraphics[width=2.5in]{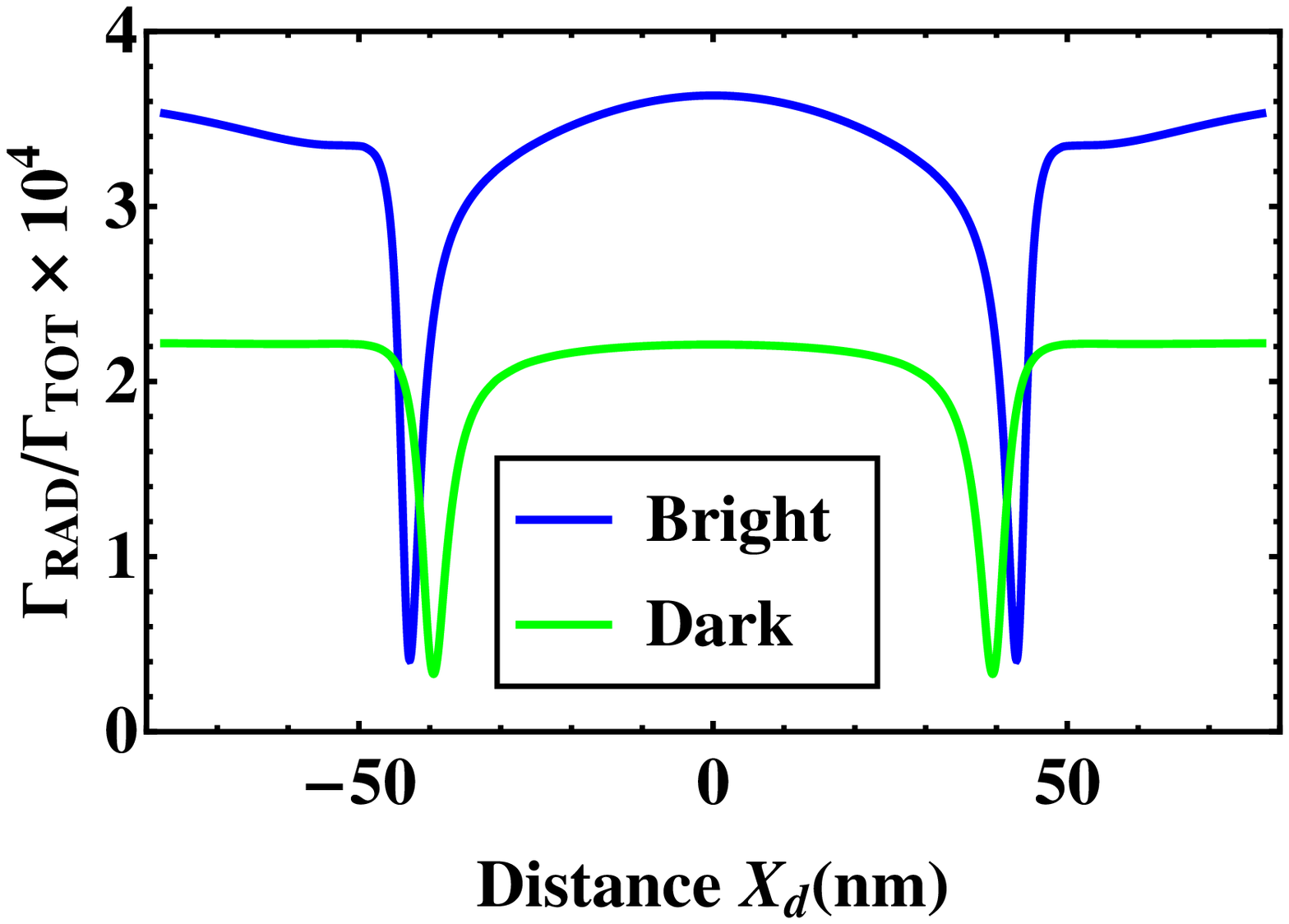}
  \caption{}\label{fig:rad_efficiency_versus_position}
\end{subfigure}%
\begin{subfigure}{.5\textwidth}
  \centering
  \includegraphics[width=2.5in]{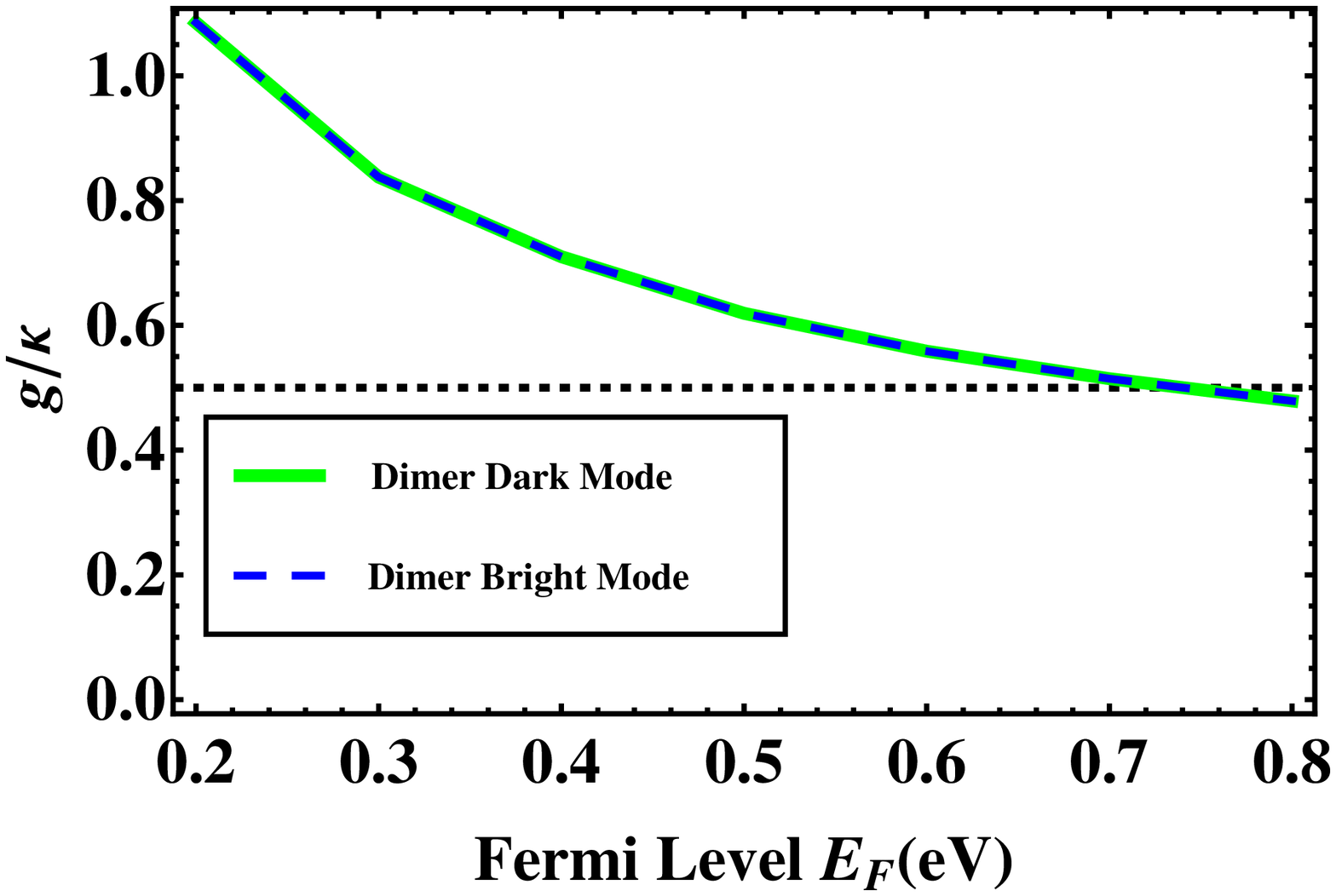}
  \caption{}\label{fig:gJCMplot}
\end{subfigure}
\caption{a) {\it Simulated (BEM) Radiative efficiency as a function of emitter position and polarization:} 
An x-polarized emitter is located at ($X_d$, 0, $Z_d$). $Z_d$ is fixed such that the emitter is located 15 nm above the closest disc. The bright mode is found to enhance radiative decay rate, but the dark mode does not. b) {\it Vacuum Rabi splitting versus Fermi level:} 
Normalized vacuum Rabi splitting for the dark and bright modes when an x-polarized emitter is located at (0, 0, $Z_d$) outside the cavity. $Z_d$ is fixed such that the emitter is located 15 nm above the closest disc. The dashed line represents $g/\kappa=1/2$ below which splitting will not be observed. ($\kappa\approx$10 meV). }
\label{fig:emitter_outside_g_by_k}
\end{figure}

We also study the dependence of the radiative efficiency, as a function of position $X_d$ of the emitter, in Fig. \ref{fig:rad_efficiency_versus_position}. Clearly, the bright mode has a higher radiative efficiency compared to the dark mode, for various locations of the emitter. We find that the radiative efficiency drops from a maximum at $X_d = 0$ nm to a local minimum, as the emitter approaches a certain horizontal distance ( $X_d = 40$ nm in the specific case of Fig. \ref{fig:rad_efficiency_versus_position}) and then rises again. This behaviour can be qualitatively explained by looking at the overlap term (not shown), which predicts that the mode excitability being maximum at $X_d = 0$ nm, drops to zero at a certain $X_d$ to subsequently rise again. Similar pattern in observed for the dark mode.

In the analysis provided here, one must note that the dominant decay pathway for the emitter in this geometry is still non-radiative. Hence the overall quantum efficiency when the emitter couples to either the dark or the bright mode, is rather small for both cases. 

\underline{Strong coupling regime}: 
Strong coupling regime in a system composed of a single graphene disc and a quantum emitter was predicted in \cite{doi:10.1021/nl201771h}. Here we briefly mention that the same can be obtained in the dimer system as well. 
However, so far we have not carried out a complete study in this direction.

In our classical electrodynamic simulations carried out for the graphene nanodisc cavity, we obtain $\Gamma_{tot}\approx\Gamma_{plasmon}>\kappa$ on resonance. This condition ensures the existence of coherent coupling between the plasmon and the emitter and the possibility of observing vacuum Rabi splitting \cite{10.1103/PhysRevB.87.115419}.

Using the approach mentioned in the Methods section, we calculated $g$ factors for different Fermi-level energies, ranging from $0.2-0.8$ eV, for different positions of the emitter both inside and outside the cavity. Our calculations for the normalized Rabi splitting are shown in Fig.~\ref{fig:gJCMplot}, for an example case of the emitter placed outside the cavity (same geometr	y as considered earlier in this section). We find that for a range of values of the doping level, we do obtain $g/\kappa > 1/2$ and we predict Rabi splitting values $g$ of up to $10$ meV at room temperature. Here for calculating the absolute value of $g$, we have used $\Gamma_0\approx 5\times10^7 s^{-1}$ \cite{doi:10.1021/nl201771h}. We note that it should possible to increase the vacuum Rabi splitting by using higher quality graphene samples, so that $\tau$ is larger. Another important trend suggested by Fig.~\ref{fig:gJCMplot} is that the value of $g/\kappa$ decreases with increasing $E_F$. This can be qualitatively understood as resulting from the increasing leakiness of the plasmon mode as the carrier concentration is increased. This results in a larger mode volume $V$, which is expected to cause a decrease in the decay rate $\Gamma_{tot}$. Since $\kappa\sim 1/\tau$ is almost independent of Fermi Level $E_F$, therefore $g/\kappa\approx\sqrt{\Gamma_{tot}/\kappa}$ decreases with increasing $E_F$.
\section{Conclusion}
We have performed a complete study of the plasmon modes of a vertically stacked dimer of graphene discs. The eigenmodes and resonant frequencies of the modes were calculated semi-analytically using a quasistatic approximation. We showed a convincing match between full wave BEM simulations and the quasistatic approach. Secondly we explained the position and polarization dependence of the LDOS using the framework of eigen-response theory. In this case also, results were consistent with simulations.

Subsequently, we focused on the dark and bright plasmon modes formed out of dipolar modes of each constituent disc of vertically stacked graphene disc dimer cavity. Due to the different symmetry of these two modes, completely different behaviour is observed in the far field response as well as decay rates. We have shown that it is possible to engineer the decay rates of a quantum emitter placed inside and near this cavity, using Fermi level tuning, via gate voltages and variation of emitter location and polarization. We highlighted that by coupling to the bright plasmon mode, the radiative efficiency of the emitter can be enhanced compared to the single graphene disc case, whereas the dark plasmon mode suppresses the radiative efficiency. Such a system can offer new degrees of freedom for controlling radiative and non-radiative emission properties of quantum emitters.

\section*{Acknowledgments}
A.K. and N.X.F. acknowledge the financial support by the NSF (grant CMMI-1120724) and AFOSR
MURI (Award No. FA9550-12-1-0488). K.H.F. acknowledges financial support from Hong Kong RGC
grant 509813. This work was performed in part at the Center for Nanoscale Systems (CNS), a member of
the National Nanotechnology Infrastructure Network (NNIN), which is supported by the National Science
Foundation under NSF award no. ECS-0335765. CNS is part of Harvard University. A.K. acknowledges Dr. Weihua Wang from the Technical University of Denmark for providing a calculation check for the single disc case.
\end{document}